\documentclass{SciPost}

\newcommand{\be}{\begin{equation}}
\newcommand{\ee}{\end{equation}}
\newcommand{\lb}{\label}

\def\backgroundMetric{\hat{g}}

\hypersetup{
    colorlinks,
    linkcolor={red!50!black},
    citecolor={blue!50!black},
    urlcolor={blue!80!black}
}

\usepackage[bitstream-charter]{mathdesign}
\DeclareSymbolFont{usualmathcal}{OMS}{cmsy}{m}{n}
\DeclareSymbolFontAlphabet{\mathcal}{usualmathcal}

\fancypagestyle{SPstyle}{
\fancyhf{}
\lhead{\colorbox{scipostblue}{\bf \color{white} ~SciPost Physics }}
\rhead{{\bf \color{scipostdeepblue} ~Submission }}

\fancyfoot[C]{\textbf{\thepage}}
}

\begin{document}

\pagestyle{SPstyle}

\begin{center}{\Large \textbf{\color{scipostdeepblue}{
Microscopic entropy of de Sitter spacetime and entropic solution to the old cosmological constant problem\\
}}}\end{center}

\begin{center}\textbf{
Mariano Cadoni\textsuperscript{1,2$\star$},
Lorenzo Herres\textsuperscript{1,2$\dagger$},
Lorenzo Orlando\textsuperscript{1,2$\ddagger$} and
Mirko Pitzalis\textsuperscript{1,2$\mathsection$}
}\end{center}

\begin{center}
{\bf 1} Dipartimento di Fisica, Universit\`a di Cagliari, Cittadella Universitaria, 09042 Monserrato, Italy
\\
{\bf 2} INFN, Sezione di Cagliari, Cittadella Universitaria, 09042 Monserrato, Italy
\\[\baselineskip]
$\star$ \href{mailto:mariano.cadoni@ca.infn.it}{\small mariano.cadoni@ca.infn.it}\,,
$\dagger$ \href{mailto:lorenzo.herres@ca.infn.it}{\small lorenzo.herres@ca.infn.it}\,,\\
$\ddagger$ \href{mailto:lorenzo.orlando@ca.infn.it}{\small lorenzo.orlando@ca.infn.it}\,,
$\mathsection$ \href{mailto:mirko.pitzalis@ca.infn.it}{\small mirko.pitzalis@ca.infn.it}\,.
\end{center}

\section*{\color{scipostdeepblue}{Abstract}}
\textbf{\boldmath{%
We study the role of Weyl symmetry breaking in conformal gravity  and the residual scale symmetry of  Einstein gravity. The corresponding action is characterized by a dimensionless coupling $\alpha$, determined by the ratio between the de Sitter and Planck scales. We show that this quantity admits a natural interpretation as the Bekenstein-Hawking entropy of de Sitter spacetime. Combining ideas from the functional renormalization group, holography, and emergent gravity, we propose a microscopic interpretation of $\alpha$ as a measure of the degrees of freedom associated with the de Sitter horizon. In this framework, the renormalization group flow of $\alpha(k)$ encodes the scale dependence of these microscopic degrees of freedom. Requiring this flow to be monotonically increasing toward the infrared leads to a cosmological constant of the same order as the observed one, suggesting an entropic solution to the old cosmological constant problem. This remarkably small value can therefore be understood as a direct consequence of the extraordinarily large number of degrees of freedom in our de Sitter universe.
}}

\vspace{\baselineskip}



\vspace{10pt}
\noindent\rule{\textwidth}{1pt}
\tableofcontents
\noindent\rule{\textwidth}{1pt}
\vspace{10pt}

\section{Introduction}
The  cosmological constant problem  remains one of the most persistent and conceptually challenging issues in modern physics, despite decades of intense investigation \cite{Weinberg:1988cp,Carroll:2000fy,Peebles:2002gy,Nobbenhuis:2004wn}. It lies at the intersection of quantum field theory, gravitation, and cosmology, and is fundamentally tied to the gravitational properties of the quantum vacuum.\\
The possibility that vacuum fluctuations of quantum fields may contribute to spacetime curvature has long been recognized. This idea naturally emerges when attempting to go beyond the Standard Model of particle physics and incorporate gravitational effects at the quantum level. The relevance of this question has been dramatically reinforced by the observational discovery that the expansion of the Universe is currently accelerating. Within the standard cosmological framework, this acceleration is attributed to a cosmological constant $\Lambda$ in Einstein's equation. 
This result, originally established through Type Ia supernovae observations \cite{SupernovaCosmologyProject:1998vns,SupernovaSearchTeam:1998fmf,Brout:2022vxf}, has been confirmed and refined by a variety of independent probes including measurements of the cosmic microwave background \cite{Planck:2018vyg} and large-scale structure \cite{SDSS:2005xqv,BOSS:2016wmc}. More recently, high precision spectroscopic surveys such as the Dark Energy Spectroscopic Instrument (DESI) have significantly improved constraints on the expansion history of the Universe. In particular, DESI data provide some of the most accurate measurements of baryon acoustic oscillations to date, allowing for possible deviations from a pure cosmological constant \cite{DESI:2024mwx,DESI:2025zgx}.\\
Traditionally, the problem is formulated in terms of a striking discrepancy between theoretical expectations and observations. A na\"ive estimate of the vacuum energy density, obtained by summing zero-point energies of quantum fields up to the Planck scale, exceeds the observed value of the cosmological constant by roughly $120$ orders of magnitude. This enormous mismatch constitutes the so called \textit{old cosmological constant problem} \cite{Weinberg:1988cp,Carroll:2000fy}. Although more refined analyses suggest that this estimate may be overstated \cite{Koksma:2011cq}, a severe hierarchy problem nevertheless persists.\\
A complementary formulation, often referred to as the \textit{new cosmological constant problem} \cite{Weinberg:2000yb}, concerns the apparent coincidence that the energy densities of matter and dark energy are of the same order of magnitude in the present epoch. Since the matter density dilutes with the expansion of the Universe while $\Lambda$ remains constant, their relative contributions evolve differently over cosmic time. Explaining why we happen to observe them at comparable scales today represents a further conceptual challenge.\\
A natural way to produce a hierarchically small value of the cosmological constant  $\Lambda$  is to use softly broken conformal/scale symmetries (see e.g. \cite{Cadoni:2006ww}) \footnote{Alternatively, in the pre-geometric gravity framework, the same result was obtained via SSB of an $SO(1,4)$ gauge symmetry \cite{Addazi:2026kam}.}. However, this is not quite straightforward for several reasons. In its usual form Einstein gravity is not conformally invariant. Even if one could manage to rewrite the classical Einstein-Hilbert  action in a scale-invariant form (like in \cite{Cadoni:2006ww}), quantum corrections will typically  break this symmetry so that there is no guarantee that the flow of $\Lambda$ in the infrared can produce  its  observed value.  On the other hand a microscopic interpretation of the Bekenstein-Hawking  entropy $S_{dS}$ associated with the de Sitter horizon strongly indicates an entropic/informational origin  of the cosmological constant problem.  In fact we have
\begin{equation}
\label{lds}
\Lambda=\frac{3\pi}{L^2_PS_{dS}},
\end{equation}
where $L_P$ is the Planck length, suggesting that the small value of $\Lambda$ may be directly related to the large entropy associated with the de Sitter horizon. Related ideas have been developed within a modular regularization of quantum phase space \cite{Freidel:2023ytq,Freidel:2022ryr}. In this framework, the microscopic ground-state degeneracy is identified with the gravitational entropy of the cosmological horizon, leading to a UV/IR relation that connects the suppression of the vacuum energy to the large size of the Universe.


In this work, we focus on the old cosmological constant problem and propose a possible entropic resolution within a framework which uses ideas coming from emergent gravity, conformal gravity,  the functional renormalization group (FRG) approach to gravity  and from the AdS/CFT correspondence. In the emergent gravity paradigm, the classical spacetime description provided by General Relativity (GR) is interpreted as an effective, coarse-grained description of underlying microscopic degrees of freedom (DOF) associated with a fundamental theory of quantum gravity. From this viewpoint, spacetime geometry is not fundamental but emerges from a more primitive, pre-geometric phase, and standard notions such as diffeomorphism invariance arise only at the macroscopic level \cite{Reuter:2008qx,Addazi:2024rzo,Capozziello:2026pys}.
This perspective suggests that the continuum description of spacetime cannot be obtained through a simple classical limit of a quantum theory defined on a fixed background. Rather, one expects a phase transition from a pre-geometric regime to the familiar geometric spacetime, in which gravity emerges as a collective phenomenon. A recent thermogravity construction has explored a different implementation of this emergent perspective, in which non-degenerate thermodynamic cycles generate controlled violations of local Lorentz invariance and energy-momentum conservation, with possible implications for cosmic acceleration and the cosmological constant problem \cite{Isichei:2025ssf}.\\
While the precise nature of the underlying microscopic theory remains unknown, there are several indications that gravity may exhibit a Weyl-invariant (conformal) phase at very high energies. If this is the case, our Universe must correspond to a phase in which Weyl symmetry is spontaneously broken, giving rise, at low energies, to Einstein gravity with a cosmological constant. Within this framework, the de Sitter (dS) length scale $L_0$ associated with the cosmological constant can be understood as emerging from the fundamental Planck scale physics, and is selected through the symmetry-breaking mechanism \cite{Cadoni:2025cmf}.\\
Weyl conformal symmetry is a  symmetry under local rescaling of the metric. Its breaking is expected to leave behind a residual symmetry, naturally identified with global scale invariance. Indeed, any theory invariant under Weyl transformations is also invariant under global rescalings. This residual symmetry plays a crucial role in gravity \cite{deAlfaro:1979hg,deAlfaro:1980cs}, as it allows Einstein gravity with cosmological constant to be reformulated in a scale-invariant manner. In this formulation, physical meaning is attached not to individual dimensionful constants, but to a specific dimensionless ratio between them:  $\alpha\propto (L_0/L_P)^2$ \cite{Cadoni:2006ww}.\\
A natural interpretation of $\alpha$ is provided by the emergent nature of Bekenstein–Hawking entropy  associated with the dS horizon. Despite significant progress, a complete microscopic understanding of the dS horizon remains elusive, with approaches ranging from string theory and the dS/CFT correspondence \cite{Strominger:2001pn} to information theory \cite{Ryu:2006bv}. In particular, the  (A)dS/CFT correspondence  \cite{Strominger:1996sh,Maldacena:1997re,Witten:1998qj} motivates the interpretation of $\alpha$ as the central charge of a dual CFT \cite{Maldacena:1997re, Susskind:1998dq}.\\
However, the emergent gravity, the (A)dS/CFT interpretation  of $\alpha$ and the conformal symmetry of gravity alone are not sufficient to give a complete and satisfactory understanding  of $\alpha$,  in particular of its quantum nature.  This is mainly because $\alpha$ will depend on the energy scale $k$ we are considering, i.e. $\alpha=\alpha(k)$.   

In this paper, we develop a microscopic explanation of dS entropy within a Wilsonian approach to gravity, which will enable us to determine $\alpha(k)$ within some  appropriate approximation.
We argue that the number of microscopic degrees of freedom associated with the dS horizon can be identified using the residual scale symmetry discussed above. Within this framework, we show how the hierarchy problem can be addressed by relating the cosmological constant to this underlying microscopic structure. We then study the scale dependence of these degrees of freedom using the Functional Renormalization Group (FRG) approach \cite{Reuter:1996cp} and demonstrate that their monotonic behaviour - motivated by its entropic interpretation and by the $C$-theorem in CFTs \cite{Zamolodchikov:1986gt} - provides an entropic solution to the old cosmological constant problem.\\
The paper is organized as follows. In \cref{Sect:Weyl_ScaleSymmetry} we discuss the breaking of Weyl symmetry and the emergence of a residual scale symmetry in Einstein gravity with a cosmological constant. We show that the theory can be rewritten in terms of a single dimensionless coupling $\alpha$ which is directly related to the Bekenstein-Hawking entropy of dS spacetime. In \cref{Sect:MicroscopicAlpha} we provide a microscopic interpretation of $\alpha$ as a measure of the degrees of freedom associated with the dS horizon, drawing on holographic and emergent gravity arguments. In \cref{Sect:FRG_Quantum_dS} we compute the running of $\alpha$ in the Wilsonian FRG approach. In \cref{Sect:Entropic_Solution} we argue that imposing the monotonicity of $\alpha(k)$, as required by its entropic and CFT interpretation, fixes the infrared value of the cosmological constant to be of the order of the observed one. Finally, in \cref{Sect:Conclusions} we draw our conclusions.

\section{Weyl symmetry of conformal gravity and the scale symmetry of Einstein gravity}
{\label{Sect:Weyl_ScaleSymmetry}}

There are several indications that conformal symmetry could play a crucial role in any UV completion of Einstein's theory of gravity \cite{Reuter:2008qx,tHooft:2010xlr,tHooft:2010mvw,Mannheim:2011ds,Rachwal:2018gwu,Giacometti:2024qva} \footnote{Throughout the paper, we use natural units $\hbar = c = 1$.}. It emerges naturally in many approaches to quantum gravity,  can be used in cosmology \cite{Penrose:2006zz} and to solve the singularity problems of GR
\cite{Modesto:2016max}.  The simplest formulation of conformal gravity uses a non minimally coupled  scalar (the dilaton) $\Phi$ and the action
\be\lb{action}
S_G=\frac{1}{16\pi}\int d^4x\,\sqrt{-g} \left[\Phi^2  R+6 \left(\partial\Phi\right)^2-2 \lambda_0 \Phi^4\right],
\ee
where $\lambda_0$ is a dimensionless, positive \footnote{In this paper we consider only the case of a positive cosmological constant. } constant. The action \cref{action} is invariant under the coordinate dependent   Weyl rescaling of the metric and of the dilaton
\be\lb{wr}
g_{\mu \nu}\to g'_{\mu \nu}= \Omega(x)^2 g_{\mu \nu},\quad \Phi\to \Phi'= \Omega(x)^{-1} \Phi.
\ee
At this level, the dilaton plays the role of a pure gauge degree of freedom. At low energy our world is not conformally invariant, thus the Weyl symmetry must be broken. The simplest and most natural way is through a spontaneous breaking triggered by a non vanishing VEV of the dilaton: 
\be\lb{vev}
\langle{\Phi}\rangle=\frac{1}{\ell}.
\ee
The length $\ell$ can be dealt with in two different ways, which correspond to the two different interpretations of the dilaton  $\Phi$. 
If one considers $\Phi^2$ as the (inverse of) coordinate-dependent Newton constant one has $\ell= L_P$, where  $L_P= \sqrt{G_0}$  is the Planck length. Introducing the cosmological constant $\Lambda_0$, we can now define the dimensionless coupling
\be\lb{alpha}
\alpha_0=\frac{1}{G_0\Lambda_0}= \frac{1}{3}\frac{L^2_0}{L_P^2},
\ee
where we have introduced the de Sitter length $L_0^2=3/\Lambda_0$.
By setting $\lambda_0= 1/\alpha_0$, \cref{action} now becomes the Einstein-Hilbert (EH) action with cosmological constant,
\be\lb{EHaction}
S_G=\frac{1}{16\pi G_0}\int d^4x\,\sqrt{-g} \left[R -2 \Lambda_0 \right],
\ee
Notice that the conformal symmetry of the action \cref{action} is completely broken as testified by the presence in EH \cref{EHaction} of the two  coupling constants $G_0,\Lambda_0$ with length dimensions  $2$ and $-2$ respectively. 

Alternatively, one can see $\Phi$, in the original spirit of Weyl, as a field characterizing the different local ``gauges''  used to measure distances and times \cite{Cadoni:2025cmf}. In this case $\ell$ gives only a  global,  still undetermined,  unit of measure. After the breaking of the Weyl symmetry, one expects the action to have  some  residual scale symmetry and the corresponding action to depend on dimensionless coupling constants only.  This is indeed the case as one can easily show directly  from the EH action \cref{EHaction}.  Using the rescaling  of the metric (see \cite{Cadoni:2006ww})
\be\lb{resc}
g_{\mu\nu}\to L_0^2  g_{\mu\nu},
\ee
the EH action  \cref{EHaction} becomes
\be\lb{EHaction1}
S_G=\frac{3 }{16\pi}\alpha_0\int d^4x\,\sqrt{-g} \left[R -6 \right].
\ee
Written in this form,  the gravitational action  is parametrized  only by the  same dimensionless coupling $\alpha_0=\lambda_0^{-1}$ as the Weyl-invariant action  \cref{action}. Moreover,  the coupling depends only on the ratio $L_0^2/L_P^2$, expressing the freedom of choosing $\ell$  arbitrarily. Notice that, being $\alpha_0^{-1}=G_0\Lambda_0$,  high (low) values of $\alpha_0$ correspond to weak (strong) couplings.

The rescaled form of the EH action \cref{EHaction1} and the related scale symmetry has long been known \cite{deAlfaro:1979hg,deAlfaro:1980cs,Cadoni:2006ww}. It has been observed that the redefinition of the metric tensor \cref{resc} renders the group dimension $d$ of the metric tensor equal to its engineering dimension   ($d=-2$) \cite{Cadoni:2006ww}. With this canonical definition of the metric field,  the action \cref{EHaction1} is invariant under  the scale transformation
$x^\mu\to \omega x^\mu,\,\,g_{\mu\nu}\to \omega^{-2} g_{\mu\nu} $.

\subsection{The coupling constant 
\texorpdfstring{$\alpha_0$}{alpha0} 
as the entropy of the de Sitter horizon}
Apart from the constant rescaling of the metric (which can be absorbed by a rescaling of the  spacetime coordinates)  the classical action  \cref{EHaction1}  admits  the same solution of the GR action \cref{EHaction}. The most general solution is the Schwarzschild - de Sitter (Kottler) solution. In this paper we will not consider the presence of matter. We will therefore take the purely  de Sitter solution  and write it in the static patch,
\be
ds^2=-\left( 1-\frac{r^2}{L_0^2}\right) dt^2+\left( 1-\frac{r^2}{L_0^2}\right)^{-1} dr^2 +r^2 d\Omega_2^2 .
\ee
The cosmological dS horizon is located  at $r=L_0$ and has associated  thermal parameters, the temperature $T_{dS}$ and entropy $S_{dS}$. In particular the Bekenstein-Hawking (BH) entropy can be expressed in terms of the coupling constant $\alpha_0$:

\be\lb{bhentropy}
S_{dS}=\frac{A(L_0) }{4G_0}= 3\pi\alpha_0.
\ee
We have therefore a direct interpretation of the dimensionless  coupling constant  $\alpha_0$ appearing in the gravitational action 
\cref{action} and \cref{EHaction1} as the entropy of the dS horizon.  This  feature supports an intrinsic entropic/informational nature of the EH gravitational action, which has been  put  forward in several approaches to (quantum) gravity \cite{Jacobson:1995ab,Padmanabhan:2009vy,Verlinde:2010hp}. As we will see shortly $\alpha_0$  also allows for a more direct interpretation  in terms of the number of microscopic degrees  of freedom (DOF) $N(L_0)$ associated with the dS spacetime.
This gives an intriguing relation   between $N$ and the coupling regime of the theory: large values of $N$ correspond to weak couplings, whereas low values give the strong coupling regime.

\section{Microscopic description of 
\texorpdfstring{$\alpha$}{alpha}}{\label{Sect:MicroscopicAlpha}}

Until now  our discussion has been purely classical.  In  the classical action  \cref{EHaction1},  $\alpha_0$ plays the simple role of a (dimensionless) coupling constant.  Its informational meaning comes from the BH entropy formula \cref{bhentropy}.  Thus, the microscopic interpretation of $\alpha_0$ is closely tied with understanding the microscopic origin  of the Bekenstein-Hawking entropy of the dS spacetime (see for instance \cite{Cadoni:2018dnd}). We generically expect this to be possible only within the quantum gravity framework. This is a quite involved problem essentially because  presently we have very few hints about de Sitter quantum gravity. Some of our understanding about the quantum gravity regime comes from string theory and the related gauge theories/gravity holographic correspondence. But string theories and their low-energy limits (supergravity, branes etc.) are most naturally formulated in anti-de Sitter (AdS) backgrounds. This feature is shared by most  UV completions of general relativity (GR)  proposed so far, for which it is very hard to accommodate stable (or even metastable) dS vacua, i.e. positive energy vacua without tachyonic excitations \cite{Dine:2020vmr,Dvali:2014gua,Witten:2001kn,Maldacena:2000mw}. 

Also the other main route to quantum gravity, namely the gauge theories/gravity holographic correspondence, becomes quite problematic in the case of dS gravity. Whereas it can be consistently formulated for the AdS spacetime taking the form of the AdS/CFT correspondence, it becomes quite problematic for the dS spacetime. Owing to several difficulties (presence of a spacelike instead of timelike asymptotic boundary, lack of unitarity of the dual CFT etc.), the status of the dS/CFT correspondence remains quite controversial.

Nevertheless, at the level of effective Einstein--scalar gravity, regular configurations interpolating between different spacetime vacua can be constructed \cite{Cadoni:2023wxa}. A related semiclassical mechanism has also been proposed in which a metastable AdS configuration decays into a dS vacuum, thereby nucleating dS spacetime \cite{Cadoni:2024pjl}. 

Leaving aside the problem of the quest for a consistent UV completion of GR, one can deal with the quantum corrections to the classical action \cref{EHaction1} within a Wilsonian approach. On general grounds one expects the classical scale symmetry of the action \cref{EHaction1} to be broken by quantum corrections, generating a dependence of the coupling constants $\alpha,\Lambda$ on the renormalization scale $k$: $\alpha_0\to \alpha(k),\, \Lambda_0\to \Lambda(k)$, together with the turning on of higher powers of the curvature terms in the action.

In view of this intricate situation we will use an eclectic approach in order to both give a microscopic interpretation of the coupling constant $\alpha $ and to calculate $\alpha(k)$. We will use methods and ideas coming from different pathways. We will use the {\sl{functional renormalization group}} approach to calculate $\alpha(k)$ and ideas coming from the {\sl{(A)dS/CFT correspondence}} and from {\sl{emergent gravity}} to give a microscopic interpretation of $\alpha(k)$.

In the AdS/CFT framework it is known that the number of quantum mechanical degrees of freedom (DOF) associated with an AdS region of size $L_0$ (the AdS radius) is given by the central charge $\mathcal{C}(L_0)$ of the dual CFT,

\be\lb{ccharge}
\mathcal{C}(L_0)=\frac{A(L_0) }{16\pi G_0}.
\ee

In three spacetime dimensions, i.e. for the  AdS$_3$/CFT$_2$ correspondence, \cref{ccharge} gives the renowned Brown-Henneaux results \cite{Brown:1986nw} \footnote{Similar results hold also in two spacetime dimensions, i.e. in the case of the AdS$_2$/CFT$_1$ correspondence \cite{Cadoni:1998sg}.}, whereas in higher spacetime dimensions $\mathcal{C}(L_0)$ is given in terms of two-point functions. 
The energy of AdS has to be understood as the total  energy of the $\mathcal{C}$ DOF and is therefore given in terms of the  negative Casimir energy of the vacuum of  the dual CFT,
\be\lb{casenergy}
E_{AdS}= -\frac{2 }{L_0}\mathcal{C}(L_0).
\ee
Owing to the present status of the dS/CFT correspondence a direct application of the above argument to the dS spacetime is not possible.
However, it is reasonable to assume that  the positive energy of the dS space,  which makes up dark energy,  takes a form similar to  \cref{casenergy}, i.e. it is obtained  from $\mathcal{N}$   excitations each of energy $E_\mathcal{N}=2/L_0$,
\be\lb{dSenergy}
E_{dS}= E_\mathcal{N} \mathcal{N}=\frac{2 }{L_0}\mathcal{N}(L_0).
\ee
On the other hand, the total energy inside the dS horizon is given by 
$E_{dS}=\rho_{dS} V_{dS}$, i.e. by
\be\lb{dSenergy1}
E_{dS}=\frac{\Lambda_0}{ 8 \pi G} \frac{4}{3} \pi L_0^3= \frac{L_0}{2G}.
\ee
Comparing \cref{dSenergy} with \cref{dSenergy1} and taking into account \cref{bhentropy} and \cref{ccharge}, one easily gets 
\be\lb{micro}
\alpha_0=\frac{4}{3}\mathcal{N}=\frac{4}{3}\mathcal{C}
\ee
implying that the number of excitations coincides with the number of DOF, $\mathcal{N}=\mathcal{C}$. This gives  a direct interpretation of the coupling constant $\alpha_0$ in terms of the number of DOF associated with the dS spacetime \footnote{A quite equivalent derivation of \cref{micro} (up to a factor of $2$) can be obtained  following Verlinde \cite{Verlinde:2016toy}, interpreting the positive energy of the dS spacetime as the excitation energy that lifts the AdS vacuum energy, that is $\mathcal{N}= 2\mathcal{C}$ and using  the Hardy-Ramanujan  formula to compute the number of ways the excitations can be distributed over the number of DOF. The factor of $2$ discrepancy  is obviously due to the different choice  of the vacuum energy. }.

It is important to stress that the identification (up to a $4/3$ factor) of $\alpha_0$ with the central charge  $\mathcal{C}$ of a dual  CFT  supports strongly   monotonicity of the growing  of $\alpha(k)$  in the renormalization group flow of the CFT  from the IR to the UV.  This  may be seen as a consequence of  Zamolodchikov's $C$-theorem in  two dimensions \cite{Zamolodchikov:1986gt}, and also of its higher dimensional generalizations \cite{Komargodski:2011vj}, although   the extrapolation of these results to the dS case needs some caution  due to the difficulties in identifying a unitary CFT dual to a dS spacetime.

In a quantum gravity framework  the coupling constant  $\alpha_0$ is expected to  depend  on the energy scale $k$  we are considering. This means that  $\alpha_0$ (hence also $L_0$ and $\mathcal{N}$) will  flow with $k$ and will be therefore promoted  to $k$-dependent  quantities:  $\alpha_0\to \alpha(k),\, L_0\to L(k), \, \mathcal{N}\to \mathcal{N}(k)$.
What is now needed is a well-defined  framework to compute these quantities.  Although   quite   useful  for  setting up the conceptual setup, the holographic (A)dS/CFT correspondence is not capable  to  give definite way to calculate $\alpha(k)$, particularly in the case of dS gravity in four spacetime dimensions. 

The functional renormalization group (FRG) approach, on the other hand,  represents a well-defined framework  in which one can calculate $\alpha(k)$, so that we will use it  in this paper.  In the FRG framework  quantum corrections are included by integrating quantum fluctuations up to the momentum scale $k^2$. The renormalized dynamics is encoded in an effective average action (EAA), which is solution of a FRG equation. By restricting our consideration to the so called  Einstein-Hilbert truncation, the EAA can be written in terms of a running Newton coupling constant $G(k)$ and a running cosmological constant $\Lambda(k)$ only, which immediately give, through \cref{alpha}, the running coupling constant $\alpha(k)$.

The description of $\alpha(k)$ as a running coupling  constant, as given in the  FRG framework,  generates  some tension with its interpretation in the holographic and emergent gravity framework, as the number of DOF associated to the dS spacetime.   In particular  the meaning of the momentum  scale  $k$  remains somehow mysterious in the latter framework. 
This difficulty is  strongly related to the problem of the localization   of the $\mathcal{N}(k)$  excitations building up the dS spacetime.  
On the one hand,  the holographic dS/CFT description  supports viewing the $\mathcal{N}(k)$  excitations  as living in a dual CFT  and $k$ as  the running  momentum scale of the dual theory. This in particular implies that $\mathcal{N}(k)$  must satisfy Zamolodchikov's $C$-theorem. On the other hand, as pointed out by Verlinde \cite{Verlinde:2016toy}, the same excitations should be considered as bulk thermal excitations responsible for the thermal entropy of the dS spacetime.
If  the $\mathcal{N}(k)$  excitations are localized in the volume of the dS spacetime the momentum scale $k$ has a  natural interpretation in terms of a subregion of size $r\le L_0$ of dS:
\be\lb{kinter}
k=\frac{\xi}{r},
\ee
where $r$ is the radial coordinate of  dS in the static parametrization and  $\xi$ is a $O(1)$ dimensionless constant. 
In this paper we will assume the validity  of this identification of the momentum scale $k$ in terms of $r$.  
The identification \cref{kinter}  together  with Zamolodchikov's $C$-theorem, which is assumed to be valid owing to the holographic meaning of $\mathcal{N}(k)$,     have an important implication  concerning the behaviour of $\mathcal{N}$  under the flow from the UV to the IR. A general feature of the    (A)dS/CFT correspondence is the UV $\to$ IR mapping  between
the bulk and the boundary CFT theory: small scales   in the CFT are mapped into large scales of the (A)dS spacetime. Thus,  the  $C$-theorem implies  that $\mathcal{N}(k)$ (hence $\alpha(k)$) when considered as a {\sl bulk quantity} must be a monotonically {\sl{decreasing}} function of $k$.  This is consistent with the na\"ive intuition we have on the localization  of the thermal excitations in dS:  larger dS regions  correspond  to a larger number of excitations.

\section{Quantum  de Sitter gravity and the functional renormalization group}{\label{Sect:FRG_Quantum_dS}}
As mentioned above, dS quantum gravity  remains  quite elusive  in most attempts to construct quantum theories of gravity. 
As long as one is simply interested in the running   of $\alpha$ in terms of the momentum scale $k$ the FRG approach is the most suitable one.  In fact, in the FRG framework $\alpha(k)$ can  be  computed  using the Einstein-Hilbert truncation, in which one only retains the couplings $G(k), \Lambda(k)$, while those corresponding to  higher curvature terms are neglected. This truncation is expected to give reliable results at large distances, the regime we mainly consider in this paper.

Renormalization group methods provide a unifying framework for understanding how physical systems behave across different length scales. Originally developed in the context of critical phenomena in statistical mechanics, these ideas have since become central also in quantum field theory and, more recently, in approaches to quantum gravity. 
Within this framework, the Wilsonian approach is certainly one of the most important, as it has been able to explain why, in the vicinity of critical points, very different systems exhibit the same universal power-law behaviour, largely independent of microscopic details.

This insight is rooted in Kadanoff's scaling hypothesis through the idea of blocking transformations. In this picture, microscopic degrees of freedom are replaced by effective variables describing the system at a coarser resolution.

The combination of blocking with an appropriate rescaling defines a Renormalization Group (RG) transformation. Repeating this transformation generates the full RG flow, along which trajectories may approach fixed points corresponding to scale-invariant theories.
The renormalization group flow suppresses irrelevant couplings, which are responsible for microscopic details, and, as a result, near criticality only a certain number of relevant directions control infrared behaviour. Thus universality classes can be understood as basins of attraction of these critical points.

A natural next step is therefore to develop a formulation capable of treating quantum field theories, where infinitely many interacting modes are present. This is precisely the goal of the Functional Renormalization Group (FRG). The central idea is to introduce an infrared cutoff scale $k$ and construct 
an effective description, in terms of a scale-dependent effective average action (EAA) $\Gamma_k$. The latter is defined similarly as the ordinary effective action $\Gamma$ but it has the additional feature of a built-in infrared cutoff at the scale $k$. Thus, quantum fluctuations with momenta $q^2>k^2$ are integrated out, while the effect of the large distance fluctuations with $q^2<k^2$ is not included in $\Gamma_k$.

The standard route for constructing $\Gamma_k$ for gravity theories is the background field method, decomposing the dynamical metric $g_{\mu\nu}$ into a fixed background $\backgroundMetric_{\mu\nu}$ and a quantum fluctuation $h_{\mu\nu}$,
\begin{equation}
    g_{\mu\nu} = \backgroundMetric_{\mu\nu} + h_{\mu\nu}.
\end{equation}
This allows to define an  EAA still invariant under background diffeomorphisms. The linear split allows to gauge fix the transformation
\begin{equation}
    \delta^{Q}\backgroundMetric_{\mu\nu} = 0\, ,\quad \delta^{Q} h_{\mu\nu} = \mathscr{L}_{v}(\backgroundMetric_{\mu\nu} + h_{\mu\nu})
\end{equation}
where $\delta^{Q}$ is the quantum gauge transformation that keeps $\backgroundMetric_{\mu\nu}$ fixed.

We then consider Einstein gravity as an effective field theory and we identify the standard Einstein-Hilbert (EH) action with the EAA $\Gamma_{k}$. In principle, the effective average action can accommodate infinite curvature terms, but the EH truncation is the simplest way to work within this framework.

The latter corresponds to the projection of the RG flow onto the 2-dimensional subspace spanned by the operators $\sqrt{g}$ and $\sqrt{g}R$. Therefore, the (Euclidean) effective average action in the Einstein-Hilbert truncation is\footnote{In this section we will  write the equations for gravity in $D$ spacetime dimensions.}
\begin{equation}
    \Gamma_k [g,\backgroundMetric] = (16 \pi G(k))^{-1} \int d^{D} x\sqrt{g} \left[- R(g)+2\Lambda(k)\right] + S_{\text{gf}}[g,\backgroundMetric] +  S_{\text{gh}}[g,c,\bar{c}],
\end{equation}
where $G(k)$ and $\Lambda(k)$ denote the running Newton constant and cosmological constant, respectively, $S_{\text{gf}}$ is the classical background gauge fixing term and $S_{\text{gh}}$ is the ghost action. In this  paper we are interested in the flow of the coupling of  ${\alpha}$ which is given by \cref{alpha} in terms of $G(k)$ and $\Lambda(k)$. In the following we will therefore only consider the EH truncation of the EAA.

Solving the EAA for Einstein-Hilbert truncation means finding the equations describing the flow of the coupling constants $G(k)$ and ${\Lambda}(k)$.
The RG flow can be formulated in a compact and exact way through the Wetterich equation \cite{Wetterich:1992yh,Reuter:1993kw,Reuter:1994sg,Reuter:1996eg,Reuter:1996ub,Reuter:2012id}. This equation describes how the scale-dependent action $\Gamma_{k}$ evolves when quantum fluctuations are integrated out progressively from high to low energies. In its general form the Wetterich equation appears as follows:
\begin{equation}{\label{Master_Equation}}
\begin{split}
    k \partial_{k} \Gamma_{k} &= \frac{1}{2} \text{STr} \left[(\Gamma_{k}^{(2)}+\mathcal{R}_{k})^{-1} k \partial_{k} \mathcal{R}_{k}\right] \,,
\end{split}
\end{equation}
where $\Gamma^{(2)}_{k}$ represents the second functional derivatives of the effective average action, corresponding to the full scale-dependent inverse propagator. The regulator $\mathcal{R}_{k}$ suppresses field fluctuations with momenta below the scale $k$, ensuring that only modes with momenta larger than $k$ are integrated out. Finally, the supertrace accounts for a trace over all field components. 

In this paper we will use an approximation of the Wetterich equation, which is quite common in the literature: the so called single-metric approximation.
In the single-metric approximation  the expectation value of the fluctuation field is set  to zero. As a consequence, all covariant derivatives are evaluated with respect to the background metric, and mixed graviton–ghost terms vanish.
The Wetterich equation then decomposes naturally into separate contributions from metric and Faddeev–Popov fluctuations. The general structure therefore can be made more explicit by separating the contributions from metric fluctuations and Fadeev-Popov ghost fields. The flow equation becomes
\begin{equation}{\label{Single_Metric_Flow_Equation}}
    k \partial_{k} \Gamma_{k}[h;\backgroundMetric, c, \bar{c}] = \frac{1}{2}\,\mathrm{Tr}\!\left[ \left(\Gamma_k^{(2)} + \mathcal{R}_k\right)^{-1}_{hh}\, \partial_t \mathcal{R}^{\,hh}_k \right]
    -\mathrm{Tr}\!\left[
    \left(\Gamma_k^{(2)} + \mathcal{R}_k\right)^{-1}_{\bar c c}\,
    \partial_t \mathcal{R}^{\,\bar c c}_k
    \right]\,.
\end{equation}
Here the first trace is computed over graviton fluctuations $h_{\mu\nu}$, while the second term is a trace over the Fadeev-Popov ghost fields $c$ and $\bar{c}$. Equation \cref{Single_Metric_Flow_Equation} is usually referred to as the single-metric flow equation \cite{Percacci:2017fkn}.

We first define dimensionless couplings
\begin{equation}
    \tilde{G} = G k^{D-2}\,,\quad \tilde{\Lambda} = \Lambda/k^{2},
\end{equation}
where $G$ and $\Lambda$ are respectively the running Newton and Cosmological constants. 
In the single-metric approximation the complete flow equation for the EH truncation reads
\cite{Percacci:2017fkn}:
\begin{equation}
\begin{split}
\label{betaf}
    \frac{d\tilde{\Lambda}}{dt} &= -2\tilde{\Lambda} + \frac{1}{2}\left(A_{1} + A_{2}\eta_{N}\right)\tilde{G} + (B_{1}+B_{2}\eta_{N})\tilde{G}\tilde{\Lambda} ,\\
    \frac{d\tilde{G}}{dt} &= (D-2)\tilde{G} + \left( B_{1} + B_{2}\eta_{N} \right) \tilde{G}^{2},
\end{split}
\end{equation}
where $t=\ln k$ is known as the renormalization group time. Regarding the remaining coefficients, $A_{1}$ and $B_{1}$ are functions of $\tilde{\Lambda}$, while $\eta_{N}$ is the anomalous dimension and depends both on $\tilde{G}$ and $\tilde{\Lambda}$. We  will first compute the dimensionless coupling $\alpha$ using the one-loop approximation and an expansion  of the coefficients  $A_1,A_2,B_1,B_2$ for small values of $\tilde \Lambda$ (\cref{sect:onellop}). Later, on \cref {sect:anapprox}, we will then  compute $\alpha$ using an analytic approximation  of the exact flow.

\subsection{One loop approximation}
\label{sect:onellop}
In this section we restrict ourselves to the one-loop approximation, which corresponds to setting the anomalous dimension $\eta_{N} = 0$ in \cref{betaf}.
Moreover, the beta functions in \cref{betaf} are usually computed using the heat kernel method.
This is equivalent to a small curvature expansion so that  the coefficients in  \cref{betaf} can be expanded for small values of  the cosmological constant only. This is a quite drastic   simplification, which holds as long as the curvature of the spacetime describing our universe is small. This is true in the cosmological regime  we are considering in this paper, i.e. a dS universe for which observations tell us that it is characterized by a very small cosmological constant. 
In the following  we  will use the solutions  of the flow equations for $\tilde{\Lambda}$  and  $\tilde{G}$ already derived  in the literature \cite{Percacci:2017fkn}. We will then   rewrite the flow equations within the given approximation in terms of  the couplings  $\{\alpha(k),\tilde{\Lambda}(k)\}$.
The flow equations for $\tilde{\Lambda},\tilde G$ thus read  
\begin{equation}
\begin{split}
\label{aflow}
    \partial_t \tilde{\Lambda} &=-2\tilde{\Lambda}+ \frac{A_1}{2}\tilde{G} + B_1 \tilde{G}\tilde\Lambda,\\
    \partial_t\tilde{G} &= (D-2)\tilde{G} + B_1 \tilde{G}^2.
\end{split}
\end{equation}

Now, we trade these two equations for the flow equations of $\tilde\Lambda$ and the dimensionless combination $\alpha \equiv
1/G\,\Lambda^{(D-2)/2} = 1/\tilde G\,\tilde\Lambda^{(D-2)/2}$.
It is not  possible to write an autonomous equation for the flow of $\alpha$, but   we can easily write the equations governing the flow of $\alpha$ and $\tilde \Lambda$:

\begin{equation}
    \partial_t \alpha = - \left(\frac{D-2}{4}\right)\frac{1}{\tilde{\Lambda}^{D/2}} \left(A_1+ \frac{2D}{D-2}B_1\tilde\Lambda\right),\label{ODE_1}
\end{equation}
\begin{equation}
    \partial_t\tilde{\Lambda}=-2\tilde{\Lambda}+ \frac{A_1}{2}\frac{1}{\alpha\tilde{\Lambda}^{(D-2)/2}} + B_1\frac{1}{\alpha \tilde\Lambda^{(D-4)/2}}.\label{ODE_2}
\end{equation}
The impossibility of writing  an autonomous flow for $\alpha$ is a direct  consequence  of the breaking of scale symmetry of the classical action \cref{EHaction1} by quantum fluctuations.

Notice also that, since $\tilde{G}$ has been traded for $\alpha$, the limit $\tilde{G} \to\infty$ and $\tilde{\Lambda} \to 0$ with $\alpha$ fixed is sensible and interesting. It should correspond to a semiclassical, large (A)dS universe with strongly coupled gravitational dynamics.

The effective average action is defined off-shell. As a consequence, intermediate quantities may depend on unphysical choices such as gauge fixing or the cutoff scheme.
On the other hand, one expects that physically relevant quantities should be well defined and unambiguous. \\
In particular, this is expected to be true for dimensionless combinations of couplings, such as $\alpha$, which should maintain a certain degree of universality, especially when one tries to extrapolate on-shell quantities.
Indeed, employing on-shell methods \cite{Benedetti_2012}, one reaches the conclusion that the beta function of $\alpha(t)$ presents no gauge dependence. Unfortunately, this dependence still enters indirectly through one of the inessential couplings, whose beta function implies a gauge-dependent running. Amusingly, though the location of fixed points need not really be observable, one also finds a robust prediction for the $\tilde{\Lambda}$ coordinate of the non-Gaussian fixed point. These features, though not restricted to the small $\tilde{\Lambda}$ expansion, are nicely illustrated in that regime.
Computing the combination \cite{Percacci:2017fkn}
\begin{equation}\label{combination}
    A_1 + \frac{2D}{D-2}\, B_1 \tilde\Lambda
= \frac{16\pi (D-3)}{(4\pi)^{D/2}\,\Gamma\!\left(\frac{D}{2}\right)}
+ \frac{8\pi \left(24 + 60D - 48D^2 - 9D^3 - 4D^4 + D^5\right)}
{(4\pi)^{D/2}\, 3D(D-1)(D-2)\,\Gamma\!\left(\frac{D}{2}\right)}\, \tilde\Lambda\:\: +\mathcal{O}(\tilde\Lambda^2) 
\end{equation}
we find, in \cref{ODE_1}, no gauge dependence to order $\tilde{\Lambda}$. 
This result will be useful to us later.

Keeping the leading terms in the small $\tilde{\Lambda}$ expansion yields the simplified evolution equations
\begin{equation}
\begin{split}
{\label{Approx_equation_1}}
  \partial_t \alpha &\approx - \left(\frac{D-2}{4}\right)\frac{A_1}{\tilde\Lambda^{D/2}},\\
    \partial_t \tilde{\Lambda} &\approx \frac{1}{2}A_1 \frac{1}{\alpha\tilde\Lambda^{(D-2)/2}}\,.
\end{split}
\end{equation}
Notice that, in dimension $D = 2$, $\alpha$ does not run. This is related to the fact that in $D=2$ gravity has no dynamical degrees of freedom and Einstein-Hilbert gravity is topological.
Taking the ratio of \cref{Approx_equation_1}, one can easily derive the equation for the trajectory $\alpha(\tilde \Lambda)$ of the flow:
\begin{equation}{\label{No_Gauge_Dependence}}
    \frac{\partial\alpha}{\partial\tilde{\Lambda}} = - \left(\frac{D-2}{2}\right) \frac{\alpha}{\tilde{\Lambda}}\, ,
\end{equation}
from which we get
\begin{equation}
    \alpha(\tilde{\Lambda}) = \alpha_0 \frac{\tilde{\Lambda}_0^{(D-2)/2}}{\tilde\Lambda^{(D-2)/2}} \, ,
\end{equation}
with $\alpha_0,\tilde\Lambda_0$ integration constants.
Notice that \cref{No_Gauge_Dependence} is gauge independent in any dimension $D$; thus, the running $\alpha(\Lambda)$ gives a reliable physical information as long as the approximation is valid. Additionally, the solution implies that $\alpha(t)\tilde{\Lambda}(t)^{(D-2)/2}=\tilde{G}(t)^{-1}$  stays constant along the flow, thus also the limit $\tilde{G}\to\infty$ is reliable.
Substituting in the equations for the running of  $\alpha$ and  $\tilde\Lambda$  one gets
\begin{equation}
\begin{split}
\label{asol}
\tilde{\Lambda}(t) &= \tilde{\Lambda}_0 + \frac{1}{2} \frac{A_1}{\alpha_0\tilde\Lambda_0^{(D-2)/2}}(t-t_0),\\
\alpha(t) & =\frac{\alpha_0\tilde{\Lambda}_0^{(D-2)/2}}{\left[\tilde{\Lambda}_0 + \frac{1}{2}\dfrac{A_1}{\alpha_0 \tilde{\Lambda}_0^{(D-2)/2}}(t-t_0) \right]^{\frac{D-2}{2}}}.
\end{split}    
\end{equation}
The coefficient $A_1$ presents no gauge dependence to order $O(\tilde\Lambda^0)$. Consequently, the running of $\alpha$ continues to give  reliable physical information within the approximation. 
In particular, in $D$ dimensions from \cref{combination}
\begin{equation}
    A_1(\tilde{\Lambda}=0) = \frac{16\pi(D-3)}{(4\pi)^{D/2} \Gamma\left(\frac{D}{2}\right)}. 
\end{equation}
The approximated flow given by \cref{aflow} describes the flow from a UV non gaussian fixed point \footnote{ Numerical  values for the dimensionless coupling constants can be found e.g  in Ref \cite{Platania:2018eka},  $g^*=0.403,\, \lambda^*=0.33$.} $\tilde G=g^*,\, \tilde\Lambda= \lambda^*$ to an IR  gaussian fixed point \cite{Percacci:2017fkn} at $\tilde G=\tilde\Lambda=0$.  On the other hand, the solutions \cref{No_Gauge_Dependence} and \cref{asol} hold in the small $\tilde{\Lambda }$ regime, i.e. near the IR fixed point. Choosing $t=t_0$ as the position of the IR fixed point, one can easily  check from \cref{asol}  that in this regime  $\tilde\Lambda$ decreases monotonically whereas the dimensionless coupling $\alpha$ increases monotonically  when the system flows towards the IR. This is fully consistent  with the microscopic interpretation of $\alpha$ discussed in  \cref{Sect:MicroscopicAlpha}. However, this monotonic behaviour seems to be an artifact of the approximation used to derive   \cref{No_Gauge_Dependence} and \cref{asol}. In fact by   considering the first subleading terms in the  small $\tilde \Lambda$ expansion one gets  the solution already quoted in   \cite{Percacci:2017fkn}, which written in terms of the momentum scale $k$, reads

\begin{align}
\label{asol1}
\tilde{G}(k) &= \frac{ G_0 k^2}{1+ \frac{ G_0 }{g^*}k^2 },\\
\alpha(k) & =\frac{\left(1+b  G_0 k^2\right)^2}{ G_0\left(a  G_0 k^4+ {\Lambda}_0 \right)}, 
\end{align}
where $a,b$ are some coefficients appearing in the  small  $\tilde\Lambda$ expansion of the coefficients $A_1,B_1$ in \cref{aflow} (see again \cite{Percacci:2017fkn}) whereas  $ G_0$ and ${\Lambda}_0$ are the IR values of $ G$ and ${\Lambda}$.
It is easy to see that $\tilde G$ is always monotonically decreasing  in the UV$\to$ IR flow,  whereas $\tilde \Lambda$ has a minimum  and $\alpha$  has a maximum at nonvanishing  values of $k$.   The complication here is due to the fact that we  do not know if this behaviour is an artifact  of the approximation or a generic feature also of the exact flow.
Another important issue  is the  dependence of the coefficients $a,b,g^*$ in  \cref{asol1} on the gauge choice. Although this gauge dependence does not invalidate  the qualitative behaviour of the coupling constants $\tilde G,\tilde \Lambda, \alpha$  it still affects the  physical outcome  of the calculation. In the next  section  we address these issues by discussing the exact flow  of the coupling constants.

\subsection{Exact flow and Analytic  approximation}
{\label{sect:anapprox}}

The  trajectories for the exact flow given by \cref{betaf} have been classified  and already  discussed at length in the literature (see e.g  \cite{Platania:2018eka}). Those of interest for us have $\tilde \Lambda>0$  and  have been classified as Type IIIa.  They originate in the  UV from a non gaussian fixed point  at $\tilde \Lambda=\lambda^*,\tilde G=g^*$ but terminate in the IR at a finite, nonvanishing  value $k=k_t$ for the cutoff scale. $k_t$  corresponds to the divergence  of the anomalous dimension $\eta$

\begin{equation}{\label{anomdim}}
  \eta=  \frac{\tilde G B_1(\tilde \Lambda)}{1-\tilde G B_2(\tilde \Lambda)} ,
\end{equation}
where $B_1(\tilde \Lambda),B_2(\tilde \Lambda) $ are some functions (see \cite{Platania:2018eka}) whose explicit form is not relevant for our purposes.

The divergence of  $\eta$  in the IR  and the appearance of the  energy scale $k_t$   is the signature of the presence of the dS horizon.
According to \cref{kinter}, the energy scale $k_t$ is related to the radius $r_H$ of the cosmological dS horizon by:

\begin{equation}{\label{dsh}}
  k_t\sim  \frac{1}{r_H}.
\end{equation}

Here, we  will use an approximate  analytic form for the RG trajectories provided in \cite{Koch:2013owa}. $\tilde G(k)$ is given as in \cref{asol1}, whereas $\tilde \Lambda(k) $ is given as a function of $\tilde G(k)$ (see \cite{Koch:2013owa,Bonanno:2024wvb}). Using these expressions one can easily write down the solutions for the RG flow $\alpha(k)$ as follows  

\begin{align}
    \frac{\alpha_*}{\alpha(k)} &= -5 + \left(5+ \frac{\alpha_*}{\alpha_0}\right)\left[1- g(k)\right]^{\frac{3}{2}} + \frac{3}{2}g(k) \left[5- g(k)\right]\label{alphaexact1},\\ 
  g(k)&= \frac{k^2}{k^2+g_0^{-1}}, \label{alphaexact2}
\end{align}

where we have used the following definitions
\begin{equation}{\label{definition}}
g(k)=\frac{\tilde G(k)}{g^*},\quad g_0=\frac{G_0}{g^*},\quad\alpha^*=\frac{1}{g^*\lambda^*}. 
\end{equation}

A nice feature  of \cref{alphaexact1} is that, differently from \cref{asol1},  we expect a very mild dependence  from the gauge choice. This is because it is expressed fully in terms of the dimensionless coupling $\alpha$, which should not have  gauge dependence, and of the rescaled gravitational coupling $g$, which by definition satisfies $0\le g\le 1$.

The behaviour of $\alpha(k)$ as given by \cref{alphaexact1} is qualitatively similar to that of the approximate solution \cref{asol1}. Starting from the UV NGFP $\alpha(k)$  grows until it reaches a maximum  at $k_M\ge k_t$ for which $g=g_M$ is solution of the equation
\begin{equation}{\label{gM}}
(5-2g)-\left(5+\dfrac{\alpha_*}{\alpha_0}\right)(1-g)^{1/2}=0.
\end{equation}
Solving this equation one finds 
\begin{equation}
    g_M= \frac{1}{8}\left[20 -\left(5+\dfrac{\alpha_*}{\alpha_0}\right)^2 +\left(5+\dfrac{\alpha_*}{\alpha_0}\right)\sqrt{\left(5+\dfrac{\alpha_*}{\alpha_0}\right)^2-24}\right]\,.
\end{equation}

\section{Entropic solution  of the cosmological constant problem}{\label{Sect:Entropic_Solution}}
In the previous sections we have seen that $\alpha(k)$ is a non-monotonic function. This seems to be not only a  property of the small $\tilde \Lambda$ expansion, but a generic feature of the exact RG  flow of Einstein-Hilbert  gravity. But this behaviour contradicts the microscopic interpretation of $\alpha$ in terms the DOF associated with the dS spacetime and with related results of the dS/CFT correspondence presented in \cref{Sect:MicroscopicAlpha}. This microscopic description requires $\alpha(k)$ to be a monotonically increasing function flowing towards the IR and reaching its absolute maximum at $k=k_t$, corresponding to the location of the cosmological dS  horizon.

On the other hand, as one can easily see from \cref{gM}, the position of the maximum of $\alpha$ depends on the  two integration constants of the  differential equations describing the flow, namely $\alpha_0, g_0$. 

Once we fix $g_0$ in terms of the Newton constant and trade $\alpha_0$ for $\Lambda_0$, the non-monotonic  behaviour of $\alpha$ becomes naturally related with the cosmological constant problem.  We have   to explain why the IR value $\Lambda_0$ of $\Lambda$ is  so tiny compared with  the $m_P^2$, or in other words, why  $\Lambda_0$ is  of the order of magnitude of the observed cosmological constant $\Lambda^{obs}\sim L^{-2}$. 
We will show now that requiring a monotonically increasing $\alpha(k)$ solves this hierarchy problem by determining the  cosmological constant  $\Lambda_0$ to be of the order of magnitude $\Lambda^{obs}$.
In order to have $\alpha(k)$ monotonically  increasing over the whole dS spacetime we must push  $k_M$ to the  dS horizon, i.e. we must require $k_t=k_M$. Parametrizing the value of $\tilde \Lambda $ at $k_M=k_t$ as $\tilde \Lambda(k_M)= \epsilon/2$ and using  \cref{alphaexact1}  with $\alpha_M^{-1}=(\epsilon/2) g_M g^* $ together with \cref{gM} we get
\begin{equation}
 g_M=\frac{\epsilon}{\lambda^*}  -1 .
\end{equation}
Being $0\le g\le 1$ we must have  $ \lambda ^*\le\epsilon\le 2\lambda ^*$: as a consequence this implies $\epsilon\sim \mathcal{O}(1)$, in agreement with the most frequently found values of $\lambda ^*$.
We can now solve \cref{alphaexact1} for $g_M(\alpha_0)$ expanding in powers of $g_M$. We find at leading order

\begin{equation}
 g_M=\frac{2 \alpha^*\lambda^*}{\alpha_0\epsilon}.
\end{equation}
Finally, using the previous equation in \cref{alphaexact2}, always at leading order in the  $g_M$ expansion,  we get
\begin{equation}
 k_t^2=k_M^2\sim \Lambda^{obs}=2\frac{\Lambda_0}{\epsilon}.
\end{equation}

Making use of \cref{alpha}  and \cref{micro} we get
\begin{equation}
  \Lambda^{obs}=\frac{3}{2 L_P^2\epsilon \mathcal{N}}.
\end{equation}
This suggests that the small observed value of the cosmological constant is a direct consequence of the enormous number of degrees of freedom contained in our de Sitter universe.

\section{Conclusion}{\label{Sect:Conclusions}}

In this paper  we have proposed an  entropic  solution  of the old cosmological constant problem.  Motivated by the relation existing in classical Einstein gravity between  the entropy of the dS spacetime and the cosmological constant  \cref{lds}  we have looked for a microscopic explanation of this formula in a quantum gravity framework. Using  ideas coming from different approaches to quantum gravity, namely conformal gravity, the emergent gravity scenario, the (A)dS/CFT correspondence and the FRG approach, we have been able to show that the IR value of a running $\Lambda$ is constrained to be of the same order of magnitude as the observed value of the cosmological constant. We have also argued  that   \cref{lds} has a microscopic interpretation  in terms of the number of the DOF $N$ associated with the dS spacetime $\Lambda\propto 1/(L_P^2 N)$. This explains the tiny value of the observed cosmological constant as  a direct consequence of the huge number of DOF building our dS universe.


Our  derivation relies on some assumptions that although well-motivated and quite commonly used  in the literature are nevertheless quite subtle.  For sake of clarity let us spell them out   explicitly again. 
We  have treated  the dS/CFT  correspondence   on the same footing as the  AdS/CFT one. In particular we have assumed   the validity in our context of  the $C-$ theorem and of the UV-IR mapping. For what concerns the FRG approach,  we have assumed the relation \cref{kinter} between the momentum scale $k$  and the radial coordinate $r$ of the dS spacetime.  This  could be criticized  because of its non-covariant nature.   However, one can easily argue that  similar results can be obtained using the  covariant identification of $k$ in terms of the proper radial distance in the dS spacetime  $\mathcal{L}_{dS}$, $k \propto 1/\mathcal{L}_{dS}$, which has been already used in the FRG derivation of quantum geometries \cite{Bonanno:2024wvb,Bonanno:2000ep,Koch:2013owa}.

Another important issue is that of the gauge-independence of the FRG equations we have used in our paper.
As  explicitly remarked  in the main body of this paper most of the  RG flow equations we have used have some hidden or explicit dependence on gauge choices, which is a quite common problem of this approach. In our paper we have tried to focus on gauge invariant expressions  whenever possible. When this was not possible we worked with quantities such as $\alpha(k)$ and $g(k)$ whose gauge dependence is expected to be mild if not completely absent.   

Finally,  our  result does not predict  a precise value of the IR value of $\Lambda_0$, which remains  an integration constant of the RG equation, as it should necessarily be. What we have found  is that  $\Lambda(k)$ must flow in the IR to the value  determined  by the size of the dS horizon. In our universe this  determines a tiny value of the cosmological constant and, correspondingly, a huge  number of DOF making up our universe. This does not  forbid   the possibility, which nature seems not to have realized, of a Planckian universe in which $\Lambda$ is of Planckian order, the size of the dS horizon is  of order Planck length and the dS universe contains only a small number of degrees of freedom.

\bibliography{Bibliography.bib}

@article{Cadoni:2025cmf,
    author = "Cadoni, Mariano and Modesto, Leonardo and Pitzalis, Mirko and Sanna, Andrea Pierfrancesco",
    title = "{Stable wormholes in conformal gravity}",
    eprint = "https://doi.org/10.48550/arXiv.2503.14214",
    archivePrefix = "arXiv",
    primaryClass = "gr-qc",
    doi = "10.1088/1475-7516/2025/06/016",
    journal = "JCAP",
    volume = "06",
    pages = "016",
    year = "2025"
}

@article{Benedetti_2012,
   title={Asymptotic safety goes on shell},
   volume={14},
   ISSN={1367-2630},
   url={http://dx.doi.org/10.1088/1367-2630/14/1/015005},
   DOI={10.1088/1367-2630/14/1/015005},
   number={1},
   journal={New Journal of Physics},
   publisher={IOP Publishing},
   author={Benedetti, Dario},
   year={2012},
   month=jan, pages={015005} }

@book{Percacci:2017fkn,
    author = "Percacci, Robert",
    title = "{An Introduction to Covariant Quantum Gravity and Asymptotic Safety}",
    doi = "10.1142/10369",
    isbn = "978-981-320-717-2, 978-981-320-719-6",
    publisher = "World Scientific",
    series = "100 Years of General Relativity",
    volume = "3",
    year = "2017"
}

@article{Koksma:2011cq,
    author = "Koksma, Jurjen F. and Prokopec, Tomislav",
    title = "{The Cosmological Constant and Lorentz Invariance of the Vacuum State}",
    eprint = "https://doi.org/10.48550/arXiv.1105.6296",
    archivePrefix = "arXiv",
    primaryClass = "gr-qc",
    month = "5",
    year = "2011"
}

@article{Strominger:2001pn,
    author = "Strominger, Andrew",
    title = "{The dS / CFT correspondence}",
    eprint = "https://doi.org/10.48550/arXiv.hep-th/0106113",
    archivePrefix = "arXiv",
    doi = "10.1088/1126-6708/2001/10/034",
    journal = "JHEP",
    volume = "10",
    pages = "034",
    year = "2001"
}

@article{Maldacena:1997re,
    author = "Maldacena, Juan Martin",
    title = "{The Large $N$ limit of superconformal field theories and supergravity}",
    eprint = "https://doi.org/10.48550/arXiv.hep-th/9711200",
    archivePrefix = "arXiv",
    reportNumber = "HUTP-97-A097, HUTP-98-A097",
    doi = "10.4310/ATMP.1998.v2.n2.a1",
    journal = "Adv. Theor. Math. Phys.",
    volume = "2",
    pages = "231--252",
    year = "1998"
}

@article{deAlfaro:1979hg,
    author = "de Alfaro, Vittorio and Fubini, S. and Furlan, G.",
    title = "{A New Approach to the Theory of Gravitation}",
    reportNumber = "CERN-TH-2799",
    doi = "10.1007/BF02729033",
    journal = "Nuovo Cim. B",
    volume = "57",
    pages = "227--252",
    year = "1980"
}

@article{deAlfaro:1980cs,
    author = "de Alfaro, Vittorio and Fubini, S. and Furlan, G.",
    title = "{Small Distance Behavior in Einstein Theory of Gravitation}",
    reportNumber = "CERN-TH-2937",
    doi = "10.1016/0370-2693(80)90548-1",
    journal = "Phys. Lett. B",
    volume = "97",
    pages = "67--72",
    year = "1980"
}

@article{Cadoni:2006ww,
    author = "Cadoni, Mariano",
    title = "{Conformal symmetry of gravity and the cosmological constant problem}",
    eprint = "https://doi.org/10.48550/arXiv.hep-th/0606274",
    archivePrefix = "arXiv",
    doi = "10.1016/j.physletb.2006.10.009",
    journal = "Phys. Lett. B",
    volume = "642",
    pages = "525--529",
    year = "2006"
}

@article{Strominger:1996sh,
    author = "Strominger, Andrew and Vafa, Cumrun",
    title = "{Microscopic origin of the Bekenstein-Hawking entropy}",
    eprint = "https://doi.org/10.48550/arXiv.hep-th/9601029",
    archivePrefix = "arXiv",
    reportNumber = "HUTP-96-A002, RU-96-01",
    doi = "10.1016/0370-2693(96)00345-0",
    journal = "Phys. Lett. B",
    volume = "379",
    pages = "99--104",
    year = "1996"
}

@article{Ryu:2006bv,
    author = "Ryu, Shinsei and Takayanagi, Tadashi",
    title = "{Holographic derivation of entanglement entropy from AdS/CFT}",
    eprint = "https://doi.org/10.48550/arXiv.hep-th/0603001",
    archivePrefix = "arXiv",
    reportNumber = "NSF-KITP-06-11, NSF-KITP-06-11",
    doi = "10.1103/PhysRevLett.96.181602",
    journal = "Phys. Rev. Lett.",
    volume = "96",
    pages = "181602",
    year = "2006"
}

@article{DESI:2025zgx,
    author = "Abdul Karim, M. and others",
    collaboration = "DESI",
    title = "{DESI DR2 results. II. Measurements of baryon acoustic oscillations and cosmological constraints}",
    eprint = "https://doi.org/10.48550/arXiv.2503.14738",
    archivePrefix = "arXiv",
    primaryClass = "astro-ph.CO",
    reportNumber = "FERMILAB-PUB-25-0169-PPD",
    doi = "10.1103/tr6y-kpc6",
    journal = "Phys. Rev. D",
    volume = "112",
    number = "8",
    pages = "083515",
    year = "2025"
}

@article{DESI:2024mwx,
    author = "Adame, A. G. and others",
    collaboration = "DESI",
    title = "{DESI 2024 VI: cosmological constraints from the measurements of baryon acoustic oscillations}",
    eprint = "https://doi.org/10.48550/arXiv.2404.03002",
    archivePrefix = "arXiv",
    primaryClass = "astro-ph.CO",
    reportNumber = "FERMILAB-PUB-24-0154-PPD",
    doi = "10.1088/1475-7516/2025/02/021",
    journal = "JCAP",
    volume = "02",
    pages = "021",
    year = "2025"
}

@article{SupernovaCosmologyProject:1998vns,
    author = "Perlmutter, S. and others",
    collaboration = "Supernova Cosmology Project",
    title = "{Measurements of $\Omega$ and $\Lambda$ from 42 High Redshift Supernovae}",
    eprint = "https://doi.org/10.48550/arXiv.astro-ph/9812133",
    archivePrefix = "arXiv",
    reportNumber = "LBNL-41801, LBL-41801",
    doi = "10.1086/307221",
    journal = "Astrophys. J.",
    volume = "517",
    pages = "565--586",
    year = "1999"
}

@article{SupernovaSearchTeam:1998fmf,
    author = "Riess, Adam G. and others",
    collaboration = "Supernova Search Team",
    title = "{Observational evidence from supernovae for an accelerating universe and a cosmological constant}",
    eprint = "https://doi.org/10.48550/arXiv.astro-ph/9805201",
    archivePrefix = "arXiv",
    doi = "10.1086/300499",
    journal = "Astron. J.",
    volume = "116",
    pages = "1009--1038",
    year = "1998"
}

@article{Brout:2022vxf,
    author = "Brout, Dillon and others",
    title = "{The Pantheon+ Analysis: Cosmological Constraints}",
    eprint = "https://doi.org/10.48550/arXiv.2202.04077",
    archivePrefix = "arXiv",
    primaryClass = "astro-ph.CO",
    doi = "10.3847/1538-4357/ac8e04",
    journal = "Astrophys. J.",
    volume = "938",
    number = "2",
    pages = "110",
    year = "2022"
}

@article{Planck:2018vyg,
    author = "Aghanim, N. and others",
    collaboration = "Planck",
    title = "{Planck 2018 results. VI. Cosmological parameters}",
    eprint = "https://doi.org/10.48550/arXiv.1807.06209",
    archivePrefix = "arXiv",
    primaryClass = "astro-ph.CO",
    doi = "10.1051/0004-6361/201833910",
    journal = "Astron. Astrophys.",
    volume = "641",
    pages = "A6",
    year = "2020",
    note = "[Erratum: Astron.Astrophys. 652, C4 (2021)]"
}

@article{SDSS:2005xqv,
    author = "Eisenstein, Daniel J. and others",
    collaboration = "SDSS",
    title = "{Detection of the Baryon Acoustic Peak in the Large-Scale Correlation Function of SDSS Luminous Red Galaxies}",
    eprint = "https://doi.org/10.48550/arXiv.astro-ph/0501171",
    archivePrefix = "arXiv",
    reportNumber = "FERMILAB-PUB-05-057-A-CD",
    doi = "10.1086/466512",
    journal = "Astrophys. J.",
    volume = "633",
    pages = "560--574",
    year = "2005"
}

@article{BOSS:2016wmc,
    author = "Alam, Shadab and others",
    collaboration = "BOSS",
    title = "{The clustering of galaxies in the completed SDSS-III Baryon Oscillation Spectroscopic Survey: cosmological analysis of the DR12 galaxy sample}",
    eprint = "https://doi.org/10.48550/arXiv.1607.03155",
    archivePrefix = "arXiv",
    primaryClass = "astro-ph.CO",
    doi = "10.1093/mnras/stx721",
    journal = "Mon. Not. Roy. Astron. Soc.",
    volume = "470",
    number = "3",
    pages = "2617--2652",
    year = "2017"
}

@article{Nobbenhuis:2004wn,
    author = "Nobbenhuis, Stefan",
    title = "{Categorizing different approaches to the cosmological constant problem}",
    eprint = "https://doi.org/10.48550/arXiv.gr-qc/0411093",
    archivePrefix = "arXiv",
    reportNumber = "ITP-UU-04-40, SPIN-04-23",
    doi = "10.1007/s10701-005-9042-8",
    journal = "Found. Phys.",
    volume = "36",
    pages = "613--680",
    year = "2006"
}

@article{Peebles:2002gy,
    author = "Peebles, P. J. E. and Ratra, Bharat",
    editor = "Hsu, Jong-Ping and Fine, D.",
    title = "{The Cosmological Constant and Dark Energy}",
    eprint = "https://doi.org/10.48550/arXiv.astro-ph/0207347",
    archivePrefix = "arXiv",
    reportNumber = "KSUPT-02-3",
    doi = "10.1103/RevModPhys.75.559",
    journal = "Rev. Mod. Phys.",
    volume = "75",
    pages = "559--606",
    year = "2003"
}

@article{Carroll:2000fy,
    author = "Carroll, Sean M.",
    title = "{The Cosmological constant}",
    eprint = "https://doi.org/10.48550/arXiv.astro-ph/0004075",
    archivePrefix = "arXiv",
    reportNumber = "EFI-2000-13",
    doi = "10.12942/lrr-2001-1",
    journal = "Living Rev. Rel.",
    volume = "4",
    pages = "1",
    year = "2001"
}

@article{Weinberg:1988cp,
    author = "Weinberg, Steven",
    editor = "Hsu, Jong-Ping and Fine, D.",
    title = "{The Cosmological Constant Problem}",
    reportNumber = "UTTG-12-88",
    doi = "10.1103/RevModPhys.61.1",
    journal = "Rev. Mod. Phys.",
    volume = "61",
    pages = "1--23",
    year = "1989"
}

@article{Verlinde:2016toy,
    author = "Verlinde, Erik P.",
    title = "{Emergent Gravity and the Dark Universe}",
    eprint = "https://doi.org/10.48550/arXiv.1611.02269",
    archivePrefix = "arXiv",
    primaryClass = "hep-th",
    doi = "10.21468/SciPostPhys.2.3.016",
    journal = "SciPost Phys.",
    volume = "2",
    number = "3",
    pages = "016",
    year = "2017"
}

@article{Susskind:1998dq,
    author = "Susskind, Leonard and Witten, Edward",
    title = "{The Holographic bound in anti-de Sitter space}",
    eprint = "https://doi.org/10.48550/arXiv.hep-th/9805114",
    archivePrefix = "arXiv",
    reportNumber = "SU-ITP-98-39, IASSNS-HEP-98-44",
    month = "5",
    year = "1998"
}

@article{Reuter:2008qx,
    author = "Reuter, Martin and Weyer, Holger",
    title = "{Conformal sector of Quantum Einstein Gravity in the local potential approximation: Non-Gaussian fixed point and a phase of unbroken diffeomorphism invariance}",
    eprint = "https://doi.org/10.48550/arXiv.0804.1475",
    archivePrefix = "arXiv",
    primaryClass = "hep-th",
    reportNumber = "MZ-TH-08-12",
    doi = "10.1103/PhysRevD.80.025001",
    journal = "Phys. Rev. D",
    volume = "80",
    pages = "025001",
    year = "2009"
}

@article{tHooft:2010xlr,
    author = "'t Hooft, Gerard",
    title = "{Probing the small distance structure of canonical quantum gravity using the conformal group}",
    eprint = "https://doi.org/10.48550/arXiv.1009.0669",
    archivePrefix = "arXiv",
    primaryClass = "gr-qc",
    reportNumber = "ITP-UU-10-30, SPIN-10-25",
    month = "9",
    year = "2010"
}

@article{tHooft:2010mvw,
    author = "'t Hooft, Gerard",
    title = "{The Conformal Constraint in Canonical Quantum Gravity}",
    eprint = "https://doi.org/10.48550/arXiv.1011.0061",
    archivePrefix = "arXiv",
    primaryClass = "gr-qc",
    month = "11",
    year = "2010"
}

@article{Mannheim:2011ds,
    author = "Mannheim, Philip D.",
    title = "{Making the Case for Conformal Gravity}",
    eprint = "https://doi.org/10.48550/arXiv.1101.2186",
    archivePrefix = "arXiv",
    primaryClass = "hep-th",
    doi = "10.1007/s10701-011-9608-6",
    journal = "Found. Phys.",
    volume = "42",
    pages = "388--420",
    year = "2012"
}

@article{Rachwal:2018gwu,
    author = "Rachwa{\l}, Les{\l}aw",
    title = "{Conformal Symmetry in Field Theory and in Quantum Gravity}",
    eprint = "https://doi.org/10.48550/arXiv.1808.10457",
    archivePrefix = "arXiv",
    primaryClass = "hep-th",
    doi = "10.3390/universe4110125",
    journal = "Universe",
    volume = "4",
    number = "11",
    pages = "125",
    year = "2018"
}

@article{Giacometti:2024qva,
    author = "Giacometti, G. and Bonanno, A. and Plumari, S. and Zappal{\`a}, D.",
    title = "{Spontaneous breaking of diffeomorphism invariance in conformally reduced quantum gravity}",
    eprint = "https://doi.org/10.48550/arXiv.2410.08916",
    archivePrefix = "arXiv",
    primaryClass = "gr-qc",
    month = "10",
    year = "2024"
}

@article{Jacobson:1995ab,
    author = "Jacobson, Ted",
    title = "{Thermodynamics of space-time: The Einstein equation of state}",
    eprint = "https://doi.org/10.48550/arXiv.gr-qc/9504004",
    archivePrefix = "arXiv",
    reportNumber = "UMDGR-95-114",
    doi = "10.1103/PhysRevLett.75.1260",
    journal = "Phys. Rev. Lett.",
    volume = "75",
    pages = "1260--1263",
    year = "1995"
}

@article{Padmanabhan:2009vy,
    author = "Padmanabhan, T.",
    title = "{Thermodynamical Aspects of Gravity: New insights}",
    eprint = "https://doi.org/10.48550/arXiv.0911.5004",
    archivePrefix = "arXiv",
    primaryClass = "gr-qc",
    doi = "10.1088/0034-4885/73/4/046901",
    journal = "Rept. Prog. Phys.",
    volume = "73",
    pages = "046901",
    year = "2010"
}

@article{Verlinde:2010hp,
    author = "Verlinde, Erik P.",
    title = "{On the Origin of Gravity and the Laws of Newton}",
    eprint = "https://doi.org/10.48550/arXiv.1001.0785",
    archivePrefix = "arXiv",
    primaryClass = "hep-th",
    doi = "10.1007/JHEP04(2011)029",
    journal = "JHEP",
    volume = "04",
    pages = "029",
    year = "2011"
}

@article{Maldacena:2000mw,
    author = "Maldacena, Juan Martin and Nunez, Carlos",
    editor = "Duff, Michael J. and Liu, J. T. and Lu, J.",
    title = "{Supergravity description of field theories on curved manifolds and a no go theorem}",
    eprint = "https://doi.org/10.48550/arXiv.hep-th/0007018",
    archivePrefix = "arXiv",
    doi = "10.1142/S0217751X01003937",
    journal = "Int. J. Mod. Phys. A",
    volume = "16",
    pages = "822--855",
    year = "2001"
}

@inproceedings{Witten:2001kn,
    author = "Witten, Edward",
    title = "{Quantum gravity in de Sitter space}",
    booktitle = "{Strings 2001: International Conference}",
    eprint = "https://doi.org/10.48550/arXiv.hep-th/0106109",
    archivePrefix = "arXiv",
    month = "6",
    year = "2001"
}

@article{Dvali:2014gua,
    author = "Dvali, Gia and Gomez, Cesar",
    title = "{Quantum Exclusion of Positive Cosmological Constant?}",
    eprint = "https://doi.org/10.48550/arXiv.1412.8077",
    archivePrefix = "arXiv",
    primaryClass = "hep-th",
    doi = "10.1002/andp.201500216",
    journal = "Annalen Phys.",
    volume = "528",
    pages = "68--73",
    year = "2016"
}

@article{Dine:2020vmr,
    author = "Dine, Michael and Law-Smith, Jamie A. P. and Sun, Shijun and Wood, Duncan and Yu, Yan",
    title = "{Obstacles to Constructing de Sitter Space in String Theory}",
    eprint = "https://doi.org/10.48550/arXiv.2008.12399",
    archivePrefix = "arXiv",
    primaryClass = "hep-th",
    reportNumber = "SCIPP 20/08",
    doi = "10.1007/JHEP02(2021)050",
    journal = "JHEP",
    volume = "02",
    pages = "050",
    year = "2021"
}

@article{Brown:1986nw,
    author = "Brown, J. David and Henneaux, M.",
    title = "{Central Charges in the Canonical Realization of Asymptotic Symmetries: An Example from Three-Dimensional Gravity}",
    doi = "10.1007/BF01211590",
    journal = "Commun. Math. Phys.",
    volume = "104",
    pages = "207--226",
    year = "1986"
}

@article{Cadoni:1998sg,
    author = "Cadoni, Mariano and Mignemi, Salvatore",
    title = "{Entropy of 2-D black holes from counting microstates}",
    eprint = "https://doi.org/10.48550/arXiv.hep-th/9810251",
    archivePrefix = "arXiv",
    reportNumber = "INFN-CA-TH-9813",
    doi = "10.1103/PhysRevD.59.081501",
    journal = "Phys. Rev. D",
    volume = "59",
    pages = "081501",
    year = "1999"
}

@article{Komargodski:2011vj,
    author = "Komargodski, Zohar and Schwimmer, Adam",
    title = "{On Renormalization Group Flows in Four Dimensions}",
    eprint = "https://doi.org/10.48550/arXiv.1107.3987",
    archivePrefix = "arXiv",
    primaryClass = "hep-th",
    doi = "10.1007/JHEP12(2011)099",
    journal = "JHEP",
    volume = "12",
    pages = "099",
    year = "2011"
}

@article{Zamolodchikov:1986gt,
    author = "Zamolodchikov, A. B.",
    title = "{Irreversibility of the Flux of the Renormalization Group in a 2D Field Theory}",
    journal = "JETP Lett.",
    volume = "43",
    pages = "730--732",
    year = "1986"
}

@article{Penrose:2006zz,
    author = "Penrose, R.",
    editor = "Prior, Christopher",
    title = "{Before the big bang: An outrageous new perspective and its implications for particle physics}",
    journal = "Conf. Proc. C",
    volume = "060626",
    pages = "2759--2767",
    year = "2006"
}

@article{Modesto:2016max,
    author = "Modesto, Leonardo and Rachwal, Leslaw",
    title = "{Finite Conformal Quantum Gravity and Nonsingular Spacetimes}",
    eprint = "https://doi.org/10.48550/arXiv.1605.04173",
    archivePrefix = "arXiv",
    primaryClass = "hep-th",
    month = "5",
    year = "2016"
}

@book{Platania:2018eka,
    author = "Platania, Alessia Benedetta",
    title = "{Asymptotically Safe Gravity}",
    doi = "10.1007/978-3-319-98794-1",
    publisher = "Springer International Publishing",
    address = "Cham",
    series = "Springer Theses",
    year = "2018"
}

@article{Bonanno:2024wvb,
    author = "Bonanno, Alfio and Cadoni, Mariano and Pitzalis, Mirko and Sanna, Andrea Pierfrancesco",
    title = "{Effective quantum spacetimes from functional renormalization group}",
    eprint = "https://doi.org/10.48550/arXiv.2410.16866",
    archivePrefix = "arXiv",
    primaryClass = "gr-qc",
    doi = "10.1103/PhysRevD.111.064031",
    journal = "Phys. Rev. D",
    volume = "111",
    number = "6",
    pages = "064031",
    year = "2025"
}

@article{Koch:2013owa,
    author = "Koch, Benjamin and Saueressig, Frank",
    title = "{Structural aspects of asymptotically safe black holes}",
    eprint = "https://doi.org/10.48550/arXiv.1306.1546",
    archivePrefix = "arXiv",
    primaryClass = "hep-th",
    doi = "10.1088/0264-9381/31/1/015006",
    journal = "Class. Quant. Grav.",
    volume = "31",
    pages = "015006",
    year = "2014"
}

@article{Bonanno:2000ep,
    author = "Bonanno, Alfio and Reuter, Martin",
    title = "{Renormalization group improved black hole space-times}",
    eprint = "https://doi.org/10.48550/arXiv.hep-th/0002196",
    archivePrefix = "arXiv",
    reportNumber = "INFN-CT-03-00, MZ-TH-00-04",
    doi = "10.1103/PhysRevD.62.043008",
    journal = "Phys. Rev. D",
    volume = "62",
    pages = "043008",
    year = "2000"
}

@article{Reuter:1996cp,
    author = "Reuter, M.",
    title = "{Nonperturbative evolution equation for quantum gravity}",
    eprint = "https://doi.org/10.48550/arXiv.hep-th/9605030",
    archivePrefix = "arXiv",
    reportNumber = "DESY-96-080",
    doi = "10.1103/PhysRevD.57.971",
    journal = "Phys. Rev. D",
    volume = "57",
    pages = "971--985",
    year = "1998"
}

@article{Wetterich:1992yh,
    author = "Wetterich, Christof",
    title = "{Exact evolution equation for the effective potential}",
    eprint = "https://doi.org/10.48550/arXiv.1710.05815",
    archivePrefix = "arXiv",
    primaryClass = "hep-th",
    reportNumber = "HD-THEP-92-61",
    doi = "10.1016/0370-2693(93)90726-X",
    journal = "Phys. Lett. B",
    volume = "301",
    pages = "90--94",
    year = "1993"
}

@inproceedings{Weinberg:2000yb,
    author = "Weinberg, Steven",
    title = "{The Cosmological constant problems}",
    booktitle = "{4th International Symposium on Sources and Detection of Dark Matter in the Universe (DM 2000)}",
    eprint = "https://doi.org/10.48550/arXiv.astro-ph/0005265",
    archivePrefix = "arXiv",
    reportNumber = "UTTG-07-00",
    doi = "10.1007/978-3-662-04587-9_2",
    pages = "18--26",
    month = "2",
    year = "2000"
}

@article{Addazi:2024rzo,
    author = "Addazi, Andrea and Capozziello, Salvatore and Marciano, Antonino and Meluccio, Giuseppe",
    title = "{Gravity from Pre-geometry}",
    eprint = "https://doi.org/10.48550/arXiv.2409.02200",
    archivePrefix = "arXiv",
    primaryClass = "hep-th",
    doi = "10.1088/1361-6382/ada767",
    journal = "Class. Quant. Grav.",
    volume = "42",
    number = "4",
    pages = "045012",
    year = "2025"
}

@article{Addazi:2026kam,
    author = "Addazi, Andrea and Meluccio, Giuseppe",
    title = "{Solution to the Cosmological Constant Problem from Pre-geometric Gravity}",
    eprint = "https://doi.org/10.48550/arXiv.2602.16840",
    archivePrefix = "arXiv",
    primaryClass = "hep-th",
    month = "2",
    year = "2026"
}

@article{Cadoni:2024pjl,
    author = "Cadoni, Mariano and Pitzalis, Mirko and Sanna, Andrea P.",
    title = "{Nucleation of de Sitter from the anti de Sitter spacetime in scalar field models}",
    eprint = "https://doi.org/10.48550/arXiv.2407.10469",
    archivePrefix = "arXiv",
    primaryClass = "gr-qc",
    doi = "10.1140/epjc/s10052-025-13960-1",
    journal = "Eur. Phys. J. C",
    volume = "85",
    number = "3",
    pages = "227",
    year = "2025"
}

@article{Cadoni:2023wxa,
    author = "Cadoni, Mariano and Oi, Mauro and Pitzalis, Mirko and Sanna, Andrea P.",
    title = "{Scalar stars and lumps with AdS or dS cores}",
    eprint = "https://doi.org/10.48550/arXiv.2311.16934",
    archivePrefix = "arXiv",
    primaryClass = "gr-qc",
    doi = "10.1103/PhysRevD.109.084031",
    journal = "Phys. Rev. D",
    volume = "109",
    number = "8",
    pages = "084031",
    year = "2024"
}

@article{Capozziello:2026pys,
    author = "Capozziello, Salvatore and Meluccio, Giuseppe",
    title = "{The Pre-geometric Origin of Geometric Trinity of Gravity}",
    eprint = "https://doi.org/10.48550/arXiv.2606.17580",
    archivePrefix = "arXiv",
    primaryClass = "gr-qc",
    month = "6",
    year = "2026"
}

@article{Reuter:1993kw,
    author = "Reuter, M. and Wetterich, C.",
    title = "{Effective average action for gauge theories and exact evolution equations}",
    reportNumber = "DESY-93-152, HD-THEP-93-40",
    doi = "10.1016/0550-3213(94)90543-6",
    journal = "Nucl. Phys. B",
    volume = "417",
    pages = "181--214",
    year = "1994"
}

@article{Reuter:1994sg,
    author = "Reuter, M. and Wetterich, C.",
    title = "{Exact evolution equation for scalar electrodynamics}",
    reportNumber = "DESY-94-017, HD-THEP-93-41",
    doi = "10.1016/0550-3213(94)90278-X",
    journal = "Nucl. Phys. B",
    volume = "427",
    pages = "291--324",
    year = "1994"
}

@article{Reuter:1996eg,
    author = "Reuter, M. and Wetterich, C.",
    title = "{Quantum Liouville field theory as solution of a flow equation}",
    eprint = "https://doi.org/10.48550/arXiv.hep-th/9605039",
    archivePrefix = "arXiv",
    reportNumber = "DESY-96-081, HD-THEP-96-11",
    doi = "10.1016/S0550-3213(97)00447-1",
    journal = "Nucl. Phys. B",
    volume = "506",
    pages = "483--520",
    year = "1997"
}

@inproceedings{Reuter:1996ub,
    author = "Reuter, M.",
    title = "{Effective average actions and nonperturbative evolution equations}",
    booktitle = "{5th Hellenic School and Workshops on Elementary Particle Physics}",
    eprint = "https://doi.org/10.48550/arXiv.hep-th/9602012",
    archivePrefix = "arXiv",
    reportNumber = "DESY-96-016",
    month = "2",
    year = "1996"
}

@article{Reuter:2012id,
    author = "Reuter, Martin and Saueressig, Frank",
    title = "{Quantum Einstein Gravity}",
    eprint = "https://doi.org/10.48550/arXiv.1202.2274",
    archivePrefix = "arXiv",
    primaryClass = "hep-th",
    doi = "10.1088/1367-2630/14/5/055022",
    journal = "New J. Phys.",
    volume = "14",
    pages = "055022",
    year = "2012"
}

@article{Witten:1998qj,
    author = "Witten, Edward",
    title = "{Anti de Sitter space and holography}",
    eprint = "https://doi.org/10.48550/arXiv.hep-th/9802150",
    archivePrefix = "arXiv",
    reportNumber = "IASSNS-HEP-98-15",
    doi = "10.4310/ATMP.1998.v2.n2.a2",
    journal = "Adv. Theor. Math. Phys.",
    volume = "2",
    pages = "253--291",
    year = "1998"
}

@article{Cadoni:2018dnd,
    author = "Cadoni, M. and Casadio, R. and Giusti, A. and Tuveri, M.",
    title = "{Emergence of a Dark Force in Corpuscular Gravity}",
    eprint = "https://doi.org/10.48550/arXiv.1801.10374",
    archivePrefix = "arXiv",
    primaryClass = "gr-qc",
    doi = "10.1103/PhysRevD.97.044047",
    journal = "Phys. Rev. D",
    volume = "97",
    number = "4",
    pages = "044047",
    year = "2018"
}

@article{Freidel:2023ytq,
    author = "Freidel, Laurent and Kowalski-Glikman, Jerzy and Leigh, Robert G. and Minic, Djordje",
    title = "{On the inevitable lightness of vacuum}",
    eprint = "https://doi.org/10.48550/arXiv.2303.17495",
    archivePrefix = "arXiv",
    primaryClass = "hep-th",
    doi = "10.1142/S021827182342004X",
    journal = "Int. J. Mod. Phys. D",
    volume = "32",
    number = "14",
    pages = "2342004",
    year = "2023"
}

@article{Freidel:2022ryr,
    author = "Freidel, Laurent and Kowalski-Glikman, Jerzy and Leigh, Robert G. and Minic, Djordje",
    title = "{Vacuum energy density and gravitational entropy}",
    eprint = "https://doi.org/10.48550/arXiv.2212.00901",
    archivePrefix = "arXiv",
    primaryClass = "hep-th",
    doi = "10.1103/PhysRevD.107.126016",
    journal = "Phys. Rev. D",
    volume = "107",
    number = "12",
    pages = "126016",
    year = "2023"
}

@article{Isichei:2025ssf,
    author = "Isichei, Raymond and Magueijo, Joao",
    title = "{Lorentz Violation in Emergent Gravity and Its Cosmological Consequences}",
    eprint = "https://doi.org/10.48550/arXiv.2511.22221",
    archivePrefix = "arXiv",
    primaryClass = "gr-qc",
    doi = "10.1103/tvmx-qk3k",
    journal = "Phys. Rev. Lett.",
    volume = "136",
    number = "23",
    pages = "231501",
    year = "2026"
}


\end{document}